# Noble-Metal-Free Photocatalytic Hydrogen Evolution Activity: The Impact of Ball Milling Anatase Nanopowders with TiH$_2$

*Xuemei Zhou, Ning Liu, Jochen Schmidt, Axel Kahnt, Andres Osvet, Stefan Romeis, Eva M. Zolnhofer, Venkata Ramana Reddy Marthala, Dirk M. Guldi, Wolfgang Peukert, Martin Hartmann, Karsten Meyer, and Patrik Schmuki\**

X. Zhou, Dr. N. Liu, Prof. P. Schmuki Department of Materials Science WW-4 LKO
University of Erlangen-Nuremberg Martensstr. 7, 91058 Erlangen, Germany E-mail: schmuki@ww.uni-erlangen.de

Dr. J. Schmidt, Dr. S. Romeis, Prof. W. Peukert Institute of Particle Technology
University of Erlangen-Nuremberg Cauerstr. 4, 91058 Erlangen, Germany.

Dr. A. Kahnt, Prof. D. M. Guldi Department of Chemistry and Pharmacy
Interdisciplinary Center for Molecular Materials (ICMM) University of Erlangen-Nuremberg
Egerlandstr. 3, 91058 Erlangen, Germany.

A. Osvet, Department of Materials Sciences 6 iMEET
University of Erlangen-Nuremberg Martensstr. 7, 91058 Erlangen, Germany.

E. M. Zolnhofer, Prof. K. Meyer Department of Chemistry and Pharmacy Inorganic and General Chemistry University of Erlangen-Nuremberg Egerlandstr. 1, 91058 Erlangen, Germany.

Dr. V. R. R. Marthala, Prof. M. Hartmann ECRC-Erlangen Catalysis Resource Center University of Erlangen-Nuremberg Egerlandstr. 3, 91058 Erlangen, Germany.

In 1972, Fujishima and Honda[1] reported photoelectrochemical water splitting into hydrogen and oxygen using a titania semiconductor single crystal as photoanode. This finding triggered immense research activities on the direct conversion of photoenergy to chemical energy (hydrogen) using suitable semiconductors.[2–5] Except for photoelectrochemical approaches to produce $H_2$ (titania as a photoelectrode), the most direct photo-catalytic way is to use semiconductor nanoparticles in suspensions of $H_2O$ – with or without sacrificial agents. In general, the two greatest challenges in using $TiO_2$ for photoelectrochemistry and photocatalysis are, on one hand, the large bandgap (allowing only for UV activation of a photocatalytic reaction) and, on the other hand, the kinetically suppressed charge car-rier transfer reactions for $H_2$ and $O_2$ generation.

To tackle the first challenge, numerous doping and bandgap engineering approaches have been explored over the past decades to shift the $TiO_2$ absorption into the visible light range of the solar spectrum by introducing suitable metal[6–8] or non-metal[9] species into the $TiO_2$ lattice. Most recently, self-doping and particularly the generation of visible light absorbing "black $TiO_2$" have attracted wide interest. To this end, a large range of reduction conditions has been investigated, namely annealing in reducing atmospheres[10–12] (Ar, Ar/$H_2$, $H_2$), cathodic reaction, chemical reduction reactions, oxidation of $Ti^{2+}/Ti^{3+}$ pre-cursor,[13] ion-bombardment,[14,15] or implantation.[16] Most of these treatments of $TiO_2$ lead to classic $Ti^{3+}/O_v$ defects, from which its black or dark-blue color and, thus, visible light absorption evolves. Such defects have also been reported to have some effect on the general photocatalytic activity of titania. [2,3,17]

The second challenge intrinsic to $TiO_2$, namely the sluggish $H_2$ generation kinetics in bias-free photocatalysis, is usually tackled by decorating $TiO_2$ nanoparticles with noble metal cocatalysts including Pt, Pd, Au, etc. These cocatalysts act as electron transfer mediators and enable "reasonable" $H_2$ evolution rates under open circuit conditions.[18–20] The use of low abundant noble metals questions, however, the economic benefit of using low-cost $TiO_2$-based catalysts for $H_2$ generation.

In a number of recent reports,[21–23] some specific modifications of $TiO_2$ are shown to substantially increase the photocatalytic activity for $H_2$ evolution without any noble-metal decoration. The formation of intrinsic defects, that is, specific configurations of $Ti^{3+}$ or vacancies ($O_v$) are responsible for the unexpected activity, as they act as cocatalytic centers in $TiO_2$ that facilitate the electron transfer to the adjacent phase.

In other words, although there is a wide range of ways to introduce $Ti^{3+}$ and $O_v$ defects into titania, it is only in a few limited cases that these states act as efficient and stable cocatalytic centers for $H_2$ evolution under open circuit conditions. Successful treatments are high pressure hydrogenation or H-ion implantation. [21,22] As such, the benefits of creating an intrinsic cocatalytic activity seem to originate from synergetic interactions between defects and hydrogen that is present during these treatments.

In this work, we demonstrate that a well-established and facile ball milling approach using mixtures of commercial anatase nanoparticles and $TiH_2$ introduces noble-metal-free photocatalytic $H_2$ activity to titania. We characterize this synergistic effect in view of the nature of defects, state of hydroxylation, and investigate the effect on the energetics and kinetics of electronic states and the resulting $H_2$ evolution performance.

Dry and wet ball milling is widely used in chemical prcessing to physically break up larger particles into smaller ones,[24,25] but it has also been demonstrated to initiate mechanochemical solid state reactions,[26–28] such as the doping of host- with guest-materials.[29–31] Via dry ball milling of micrometer-sized $TiO_2$ particles, under common parameters, the grain sizes are typically reduced to 10–30 nm with a wide particle size distribution. The process is generally accompanied with severe lattice distortion and phase transformations.[32,33] If, however, comparably uniform nanoscale (≤20 nm) $TiO_2$ material is used as starting material, the short-term effect of ball milling is the introduction of oxygen vacancies and corresponding $Ti^{3+}$ states. In this context it is important to note that for particle size of 10–30 nm, anatase has been reported by theory and experiments to be the thermodynamically stable polymorph of $TiO_2$.[13,34] This is in stark contrast to larger particles, for which rutile is thermodynamically more stable. As a result, by ball milling nanoscale anatase material for comparably short time, phase transitions to rutile may be delayed as long as particle deformation is not sufficiently severe.

In the present work, we selected a commercially available anatase nanopowder with a nominal particle size of 20 nm for the introduction of ball milling induced defects, combined with $TiH_2$ as hydrogen source during the ball milling process.

In order to assess the success of this strategy, various ball milling conditions, ratios of $TiO_2/TiH_2$, as well as various reference samples were investigated and the photocatalytic $H_2$-evolution activity was evaluated under AM1.5 (100 mW cm$^{-2}$) solar simulator illumination using suspensions of the particles in a $H_2O$/methanol solution (experimental details are given in the Supporting Information). The results of **Figure 1**a show that indeed under adequate ball milling parameters and $TiH_2$ concentrations, milled powders are obtained that provide a strong enhancement of the photocatalytic $H_2$ generation. Evidently, both ball milling duration as well as $TiH_2$ loading strongly affect the photocatalytic $H_2$ generation – both lead at a lower magnitude to a beneficial effect whereas at a higher magnitude they become detrimental. In the data of Figure 1a under otherwise same conditions, an optimum activity is obtained for powder containing 2% $TiH_2$ milled for 15 min. For this sample a photocatalytic $H_2$ evolution rate of ~220 $\mu$mol h$^{-1}$ g$^{-1}$ is achieved without the presence of any noble metal cocatalyst. This rate is comparable with the activation reached by high pressure hydrogenation treatments and/or H-ion implantation.[22] It is important to note that no significant photocatalytic $H_2$ evolution can be observed for (i) ball milled anatase alone, (ii) pure $TiH_2$, or (iii) any physical mixtures of anatase and $TiH_2$. In other words, clearly ball milling and the presence of an adequate amount of $TiH_2$ are required to achieve this beneficial effect on $H_2$ evolution.

A most obvious effect of ball milling anatase or mixtures with $TiH_2$ powders is a change of color (see Figure S1, Supporting Information). With an extended ball milling time but also with a higher $TiH_2$ concentration, increasingly grey to black samples are obtained. The corresponding light reflectance spectra (Figure S1b,c,

Supporting Information) show an accordingly light absorption increase in the visible range. Nevertheless, increasing visible light absorption is not the key origin to the enhanced $H_2$ production, as experiments performed using only visible light irradiation (≥420 nm) did not yield significant amounts of generated $H_2$ (see Figure 1a).

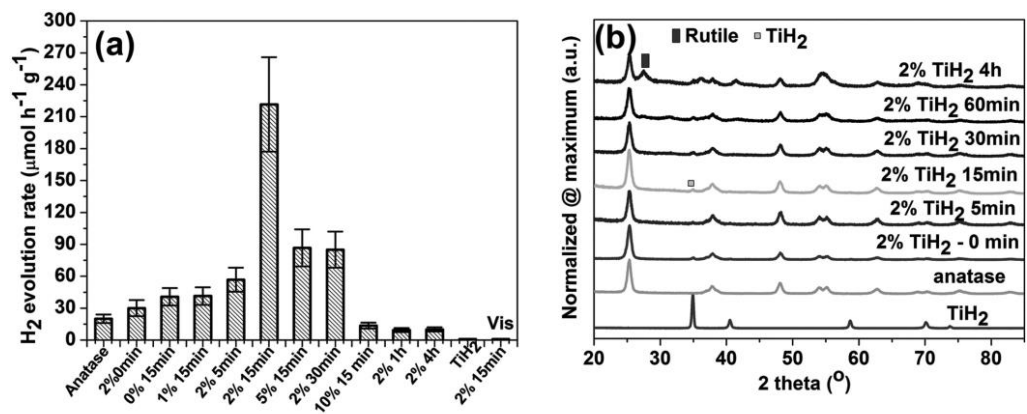

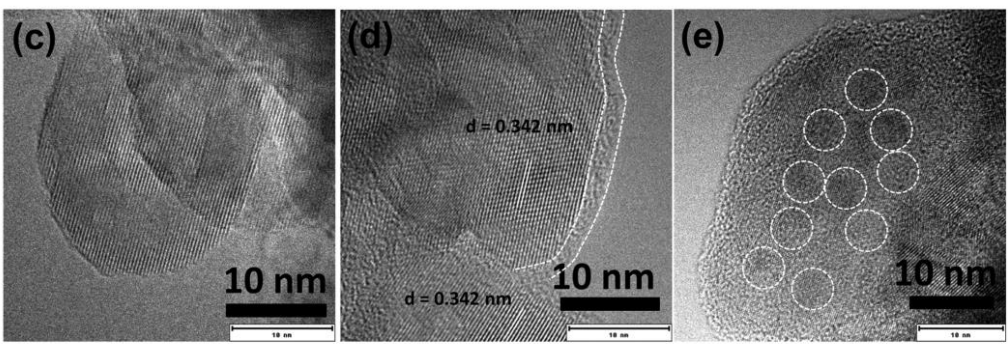

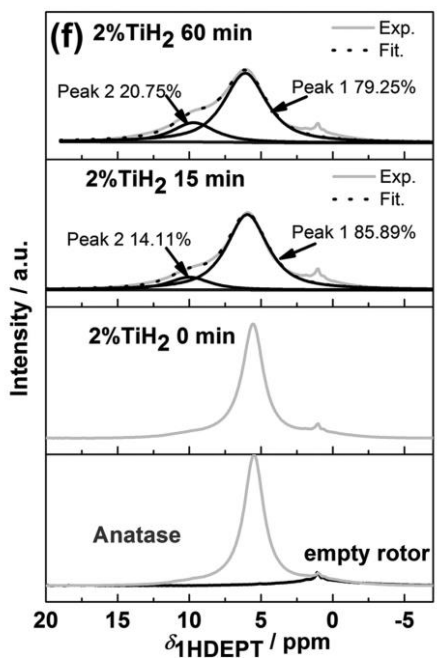

**Figure 1.** a) Photocatalytic $H_2$ evolution from $TiO_2$ anantase after ball milling process under AM1.5 (100 mW cm$^{-2}$) solar simulator illumination measured in a 50 vol% methanol–water electrolyte. b) Powder XRD patterns for samples obtained under different conditions. All the peaks are normalized to the maximum of intensity at 2 theta of 25.2°. The $TiH_2$ XRD pattern is normalized at 2 theta of 34.9°. c) TEM image of the unmilled white anatase and anatase after ball mill for d) 15 min and e) 240 min with 2% $TiH_2$, respectively. f) Solid-state $^1H$ MAS NMR spectra of representative samples.

Characterization of the powder samples in view of their crystalline structure was carried out with X-ray diffraction analysis (XRD) (Figure 1b; Figure S2, Supporting Information), high- resolution transmission electron microscopy (HRTEM) for representative samples (Figure 1c–e; Figure S4, Supporting Information), and Raman spectroscopy (**Figure 2**a,b; Figure S5, Supporting Information). Studies on the uptake and nature of hydrogen species were carried out by nuclear magnetic resonance ($^1H$ MAS NMR).

XRD (Figure 1b) shows the samples to be crystalline before and after ball milling treatments. For the main anatase peak at $2\theta=25.2°$ (PDF No. 00-021-1272, anatase/$TiO_2$, tetragonal), no change can be detected up to 1 h ball milling, i.e., in this time no obvious crystallographic changes in the powder and the mean diffraction cross-section take place. This is in line with scanning electron microscopy (SEM) in Figure S3 (Supporting Information) where clear alternations in the particle morphology only occur for milling times ≥60 min. Moreover, this finding confirms that for the nanosized material used here no further particle size reduction due to ball milling is observed. However, for samples ball milled ≥60 min, forma- tion of a clear rutile peak ($2\theta=27.4°$ PDF No. 00-001-1292, rutile/$TiO_2$, tetragonal) can be found while in spectra up to 4 h still a $TiH_2$ peak can be detected (for extended times see Figure S9, Supporting Information). SEM images for these samples (≥60 min) (Figure S3, Supporting Information) show a gradually increasing severe deformation in the nanoparticle morphology due to mechanochemical stressing. It appears that rutile can be formed once particles are sufficiently deformed. A comparison of HRTEM image for the original anatase mate- rial, the optimum ball milled sample (2% TiH$_2$, 15 min), and a sample ball milled for 4 h, shows a steadily increasing damage (Figure 1c–e). However, the main lattice fringes in the intact part of the particle remain exclusively anatase (up to 60 min ball milling) with a lattice spacing of 0.342 nm (i.e., corresponding to the XRD peak at 2$\theta$=25.2°). Interesting for the particles ball milled for 15 min with 2% TiH$_2$, a thin amorphous layer (≈2 nm) on the outside shell of particle becomes visible (Figure 1d), which is not apparent for nonmilled anatase particles (Figure 1c). This amorphous shell does obviously not sufficiently reduce the diffraction cross-section of the particles to be detected in XRD. It is, however, noteworthy that literature, for a number of reduction treatments of anatase nano- particles, reports similar amorphous layers.[19,35] For samples ball milled as long as for 4 h, the HRTEM image (Figure 1e) shows the particle edges to become less sharp and increasingly blurred but still the anatase lattice fringes can be clearly seen. For longer ball milling times, the amorphous (defective) parts in the particles increase, and finally anatase crystallites remain embedded in an amorphous matrix.

To examine the role of hydrogen in the ball milling process, we characterized key samples with $^1$H MAS NMR. Spectra for samples before and after ball milling are shown in Figure 1f. In the neat anatase spectra a signal at ≈5.58 ppm is present which is a common feature obtained from anatase powders.[22] This signal is typically ascribed to bridging TiOH surface groups.[36,37] However, for all ball milled samples, there is an additional signal at ≈10 ppm. Such high chemical shifts have in literature been attributed to strongly acidic TiOH groups pre- sent in internal interstitial positions, namely at buried phase junctions.[36,37] The results thus indicate that the amount of hydrogen located at these internal sites is increasing, as the ball milling time increases and as more defects are introduced in the particle.

In Raman spectra (Figure 2a,b; Figure S5, Supporting Infor- mation), we find that the ball milling process from 0 to 30 min with 2% TiH$_2$ shows typical five anatase Raman transitions (tetragonal space group of I41/amd, 141) with E$_g$ (144.4 cm$^{-1}$), E$_g$ (196.4 cm$^{-1}$), B$_{1g}$ (394.5 cm$^{-1}$), A$_{1g}$ (515.1 cm$^{-1}$), and E$_g$ (636.8 cm$^{-1}$). For ball milling times of 60 min or 4 h, rutile modes (tetragonal space group of P42/mnm,136) of $E_g$ (446 cm$^{-1}$) and A1g (619 cm$^{-1}$) overlapping with the anatase modes, contribute to the overall spectra. This leads to a wave-number shift in comparison to standard anatase Raman transitions and a high scattering background.[38–40] For the main Eg peak of anatase (Figure 2b), we notice a blueshift of the mode after 15 min ball milling time. In general, blueshifts and line widening are ascribed to various causes, i.e., phonon confinement effects (due to particle size reduction),[38] the presence of lattice strain,[41,42] or defects - namely $Ti^{3+}/O_v$ configuration.[40] As XRD does not indicate a particle size change for anatase particles (in the relevant size scale for phonon confinement,[38]) we ascribe the effect observed here to defect formation. The results also show that the addition of TiH$_2$ significantly contributes to the generation of defects. Samples without ball milling (2% - 0 min) or without TiH$_2$ (0% - 15 min) do not show any difference of the spectra compared with anatase powders (Figure S5, Supporting Information) and for the same ball milling time (15 min), a higher loading of TiH$_2$ leads to higher peak shift. (Please note that pure TiH$_2$ is not Raman active.)

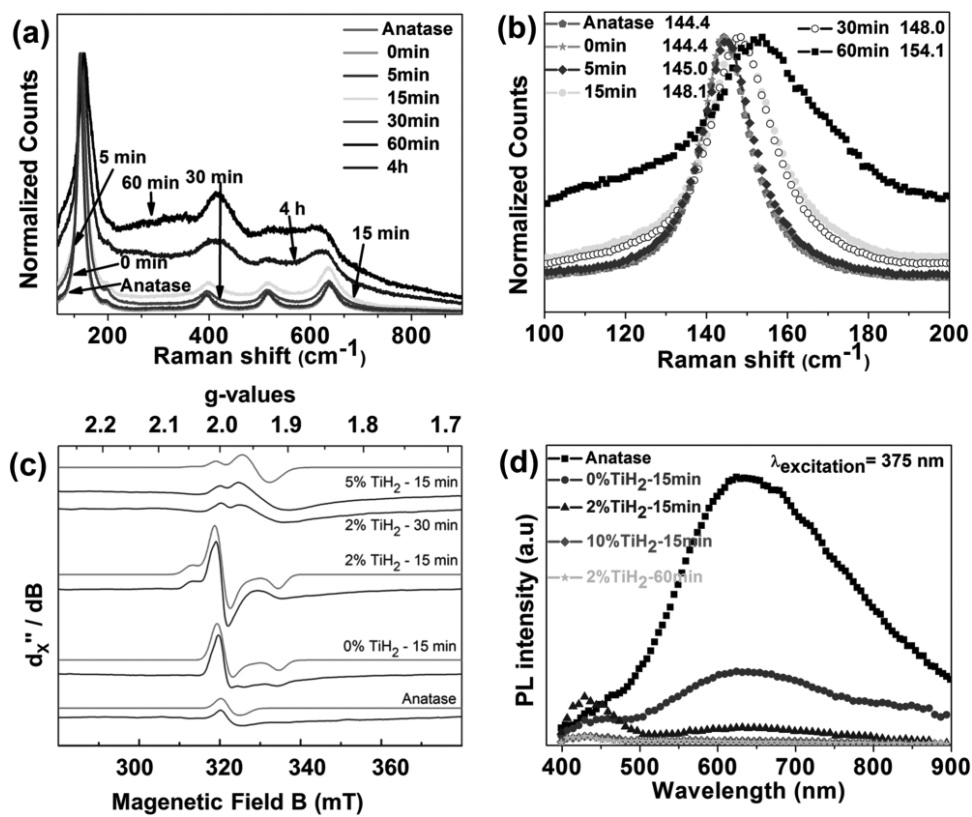

**Figure 2.** a) Raman spectra of samples with 2% TiH$_2$ ball milled after different duration (0

min to 4 h) under an excitation wavelength of 532 nm. b) Enlarged Raman spectra from 100 to 200 cm$^{-1}$. c) Solid-state EPR spectra and simulation for the anatase powder and ball milled anatase under different conditions. d) Photoluminescence spectra for representative samples under 375 nm laser excitation.

In order to further characterize the electronic nature of defects created by the ball milling process we carried out electron paramagnetic resonance (EPR) and photoluminescence (PL) measurements for various relevant samples. Figure 2c shows continuous wave (CW) X-band EPR data and simulated curves of nonmilled anatase and ball milled with TiH$_2$ (for detailed experimental conditions and simulation parameters see the Supporting Information). Pure anatase presents only a signal at $g_1$=1.998, $g_2$ =1.998, and $g_3$ =1.96 (typically assigned to Ti$^{3+}$/O$_v$ naturally present in commercial anatase). After ball milling without TiH$_2$, only subtle shifts of original signal ($g_1$ =1.998, $g_2$ = 1.998, and $g_3$ = 1.912) are observed, which may be attributed to classic ball milling lattice distortion.[43] For samples with 2% TiH$_2$, after 15 min ball milling – i.e., for the most photocatalytic active sample, an additional signal ($g_1$= 2.045, $g_2$ = 1.998, and $g_3$ = 1.998) is detected, which we assign to specific defects introduced by ball milling in the presence of TiH$_2$. Please note that pure TiH$_2$ lacks that signal. However, for longer ball milling times with TiH$_2$ and higher TiH$_2$ doses a third signal ($g_1$ = 1.946, $g_2$ = 1.946, and $g_3$ = 1.946) starts to dominate. It broadens the entire spectrum and is ascribed to Ti$^{3+}$ positions [44,45] in a distorted or amorphous lattice. Based on our experiments, the nature of these states (or their amount) is not beneficially contributing to the cocatalytic activation but rather leads to a deactivation of the material when more severe TiH$_2$ ball milling treatments are used.

More information on the energetics of different defects is derived from PL measurements. Figure 2d shows a comparison of the room temperature PL of anatase before and after different milling processes. In anatase, the PL intensity peaks at around 600–700 nm with a dropping tail toward shorter wavelength. This main emission at 600–700 nm is typical for TiO$_2$ nanoparticles[46–48] and is a superposition of trapped exciton and various defect-related emission bands in anatase nanoparticles. For pure $TiH_2$ no PL is discernable. After ball milling plain anatase, and even more so in the presence of $TiH_2$, the overall PL intensity is attenuated and continues to deintensify as a function of ball milling time or $TiH_2$ content. In fact, for the high defect- concentration samples, that is, either 2% $TiH_2$-60 min or 10% $TiH_2$-15 min, no appreciable PL is detectable – this is likely due to the dominance of nonemissive recombination (absorption) processes. In contrast, for the most active sample a new PL peak at 400–450 nm is observed which is ascribed to the presence of $Ti^{3+}$ states. These states lie at an energy of 3.1–2.95 eV (considering the onset and peak of the new PL emission at 400 nm and peaking at 420 nm, respectively) and thus around 0.2 eV below the conduction band of anatase (see also Figure S1, Supporting Information). Please note that although EPR and PL suggest the presence of $Ti^{3+}$ states. X-ray photoelectron spectroscopy (XPS) spectra fail to corroborate their existence (Figure S6, Supporting Information), as their surface concentration (penetration depth of XPS: some nm; detection limit ≈1%) is too low for detection (see the Supporting Information).

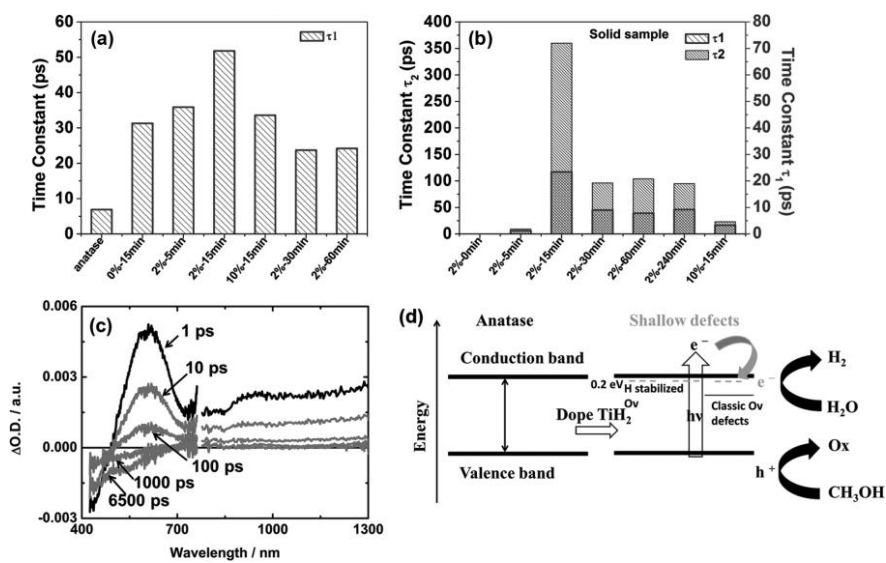

**Figure 3.** a) Plot for the time constants $\tau_1$ that are used to fit the spectra in Figure S7 (Supporting Information) as suspensions in 50% MeOH aqueous solution saturated with argon. b) Plot for the time constant $\tau_1$ and constant $\tau_2$ that are used to fit the spectra in panel (c) and Figure S8

(Supporting Information) as thin solid film. c) Differential absorption spectra (visible and near-infrared) obtained from femtosecond flash photolysis (258 nm) of 2% $TiH_2$ with $TiO_2$ anatase ball milled for 15 min as solid state at different time delays. d) Schematic drawing of charge transfer process of samples under illumination.

A main point in order to assess the role of defects in the form of localized states for photocatalysis is their effect on charge carrier lifetime and transfer kinetics. Therefore, we carried out femtosecond laser pump probe experiments in suspensions and for solid state thin film samples. **Figure 3** shows a comparison of relevant charge carrier lifetimes extracted from the differential absorption spectra for key samples (examples of spectra are given in Figure 3c). Analyses of the spectra and the associated absorption time profiles were carried out by means of global fitting as outlined in detail in the Supporting Information. First, differential absorption changes in the visible region, that is, between 400 and 600 nm, are assigned to absorption originating from holes that are formed upon photoexcitation. Second, in the near-infrared region, that is between 500 and 900 nm, the differential absorption is ascribed to trapped electrons. Finally, absorption of "free" electrons is observed throughout the entire visible and near-infrared region with a particularly strong fingerprint at ≈1200 nm. From $\tau_1$ and $\tau_2$ in 50% methanol solutions (a) and in air (b) in Figure 3 we infer for the ideal ball milled samples significantly longer charge carrier lifetimes. Lifetimes from suspension experiments better reflect the situation of aqueous solution $H_2$ evolution experiments than solid-state experiments (see also Figure S10, Supporting Information). Importantly, in both cases, the states, which are clearly detectable in EPR and PL, have a pronounced effect on the photogenerated charge carriers and their life-times. Reassuring is the fact that the lifetime maximizes for the sample that exhibits optimum photocatalytic conditions.[12,48–53]

Overall, the present work shows that ball milling nanosized titania with $TiH_2$ creates structural damages in the lattice in the form of vacancies, amorphization, and $Ti^{3+}$ states. In addition, ball milling leads to the incorporation of strongly acidic OH groups into titania. Both factors are crucial to reach a synergistic interplay for photocatalytic $H_2$ evolution, with a high degree of stability (Figure S11,

Supporting Information).

The results may thus be summarized by the scheme shown in Figure 3d. From EPR, PL, and femtosecond experiments we derive that those states, which are responsible for the $H_2$ evolution catalysis, are energetically close to the conduction band (≈0.2 eV below). Considering the lack of activity without $TiH_2$, these states are likely stabilized by lattice incorporated OH-groups.

Excessive damage by, for example, long ball milling time or high $TiH_2$ concentrations leads to considerable amorphization and, thus, to high densities of recombination sites. Also, the formation of classic $O_v$ which are situated 0.8–1.0 eV below the conduction band, impacts the excited state carrier life- times,[54] as they take over the key role in terms of charge carrier recombination.

The present study thus demonstrates a remarkable activation of commercial anatase powders for photocatalytic $H_2$ evolution after a facile ball milling treatment with $TiH_2$. The combined ball milling/$TiH_2$ treatment activates anatase nanoparticles without the use of any noble-metal cocatalyst. Overall, the sample activation is realized by creating conduction band- close states that extend the lifetimes of photogenerated electrons – which are regarded as a key factor for the observed enhanced photocatalytic $H_2$ evolution activity of $TiO_2$ anatase nanoparticles.

**Supporting Information**

Supporting Information is available from the Wiley Online Library or from the author.

**Acknowledgements**

The authors would like to acknowledge ERC, DFG, and the Erlangen DFG cluster of excellence (EAM) for the financial support. The authors would also thank Nhat Truong Nguyen for the TEM measurements and analysis.